\documentclass[a4paper,aps,prd,10pt,preprintnumbers,showpacs,twocolumn,superscriptaddress,nofootinbib,amsmath,amssymb]{revtex4-1}
\usepackage{graphicx}
\usepackage{cmap}

\def\imo{i}
\def\re#1{\mathrm{Re}(#1)}
\def\im#1{\mathrm{Im}(#1)}
\def\K{{\cal K}}
\def\Order#1{{\cal O}\left(#1\right)}

\begin{document}

\title{Long-lived Quasinormal Modes and Gray-Body Factors of black holes and wormholes in dark matter inspired Weyl Gravity}
\author{B. C. Lütfüoğlu}
\email{bekir.lutfuoglu@uhk.cz}
\affiliation{Department of Physics, Faculty of Science, University of Hradec Kralove, \\
Rokitanskeho 62/26, Hradec Kralove, 500 03, Czech Republic. }

\begin{abstract}
We calculate quasinormal modes and gray-body factors of a massive scalar field in the background of three compact objects in the Weyl gravity: Schwarzschild-like black holes, known as  Mannheim-Kazanas solution, non-Schwrazschild-like black holes and traversable wormholes found recently in [P. Jizba, K. Mudruňka, Phys.Rev.D 110 (2024) 12, 124006]. We show that the spectrum of the massive field is qualitatively different from massless one both in the frequency and time domains. While the mass term leads to much longer lifetime of the modes, the arbitrarily long-lived modes, known as quasi-resonances, are not achieved.
\end{abstract}

\maketitle

\section{Introduction}

Understanding the large-scale structure and evolution of the universe remains one of the primary challenges in modern physics. Two of the most pressing unresolved issues are the observed accelerated expansion of the universe and the nature of the dark matter that appears to dominate galactic dynamics. While conventional approaches introduce additional fields, dark matter components, or modifications to Einstein’s equations, a more fundamental perspective emerges from Weyl gravity, a purely conformal extension of general relativity. In their seminal work \cite{Mannheim:1988dj}, Mannheim and Kazanas demonstrated that Weyl gravity, without an Einstein-Hilbert term, naturally accommodates both an effective cosmological constant and galactic rotation curve corrections. Unlike phenomenological modifications of gravity, these corrections arise directly from the structure of the field equations, which are of fourth order and contain four integration constants—two of which can be interpreted as encoding cosmological and dark matter-like effects.

Beyond its implications for cosmic dynamics, Weyl gravity also provides novel black hole and wormhole solutions with distinct phenomenological properties. While standard general relativity requires exotic matter sources, such as phantom fields or finely tuned matter configurations, to support traversable wormholes \cite{Visser:1995cc,Blazquez-Salcedo:2018ipc,Azad:2022qqn,Battista:2024gud}, recent studies have demonstrated that pure Weyl gravity allows for such solutions without violating energy conditions \cite{Jizba:2024owd}. Moreover, it supports non-Schwarzschild black holes that differ qualitatively from their Einsteinian counterparts \cite{Jizba:2024owd,Mannheim:1988dj}. Given these distinctions, the study of perturbations in Weyl gravity is particularly compelling, as it provides an avenue for testing the theory against observational data and exploring the stability of its exotic solutions.

A key aspect of any compact object’s stability and observational signature is its quasinormal mode (QNM) spectrum, which governs the response of the spacetime to perturbations \cite{Nollert:1999ji, Kokkotas:1999bd, Konoplya:2011qq}. Although the boundary conditions defining QNMs are similar for black holes and wormholes \cite{Konoplya:2005et}, the resulting spectra can exhibit significant differences, potentially allowing wormholes to mimic black holes \cite{Damour:2007ap,DeSimone:2025sgu}. However, previous studies have shown that wormhole perturbation spectra also introduce distinguishing features 
and differences in late-time tails \cite{Cardoso:2016rao, Churilova:2019qph}, which could serve as observational signatures. This makes the analysis of QNMs in Weyl gravity particularly relevant, as it may reveal qualitative departures from general relativity and help assess the viability of the theory in the context of gravitational wave astronomy.

While QNMs of Schwarzschild-like Mannheim-Kazanas black holes have been examined in several works \cite{Momennia:2019cfd, Momennia:2019edt, Momennia:2018hsm, Konoplya:2020fwg, Malik:2024bmp}, these studies have primarily focused on the Schwarzschild-like branch of modes. However, as shown in \cite{Konoplya:2020fwg}, the complete perturbation evolution in this setting involves three distinct stages: an early Schwarzschild-like phase, an intermediate phase dominated by dark matter effects, and a final stage governed by an effective de Sitter geometry. A comprehensive frequency-domain analysis of these stages remains an open problem. The QNMs of massless fields in the background of the traversable wormhole and non-Schwarzschild black hole solutions \cite{Jizba:2024owd} have been studied in \cite{Konoplya:2025mvj}.

Extending this analysis to massive scalar fields introduces additional complexities and rich phenomenology, as demonstrated in previous studies on massive QNMs \cite{Konoplya:2004wg,Konoplya:2017tvu,Zhidenko:2006rs,Konoplya:2005hr,Ohashi:2004wr,Zhang:2018jgj,Aragon:2020teq,Ponglertsakul:2020ufm,Gonzalez:2022upu,Ponglertsakul:2020ufm,Burikham:2017gdm}. Massive terms in the field equations naturally arise in various contexts, such as higher-dimensional brane-world models, where the mass parameter encodes effects of the bulk on brane-localized fields \cite{Seahra:2004fg}. Additionally, in massive gravity theories, long-wavelength gravitational modes could provide insights into the nature of gravitational waves observed by pulsar timing arrays \cite{Konoplya:2023fmh, NANOGrav:2023hvm}. Notably, massive fields allow for the existence of arbitrarily long-lived QNMs, known as quasi-resonances \cite{Ohashi:2004wr,Konoplya:2004wg}, which are present across different spin fields \cite{Konoplya:2005hr,Fernandes:2021qvr,Konoplya:2017tvu,Percival:2020skc} and in a variety of compact objects, including black holes \cite{Zhidenko:2006rs,Zinhailo:2018ska,Bolokhov:2023bwm,Skvortsova:2024eqi} and wormholes \cite{Churilova:2019qph}. At the same time the arbitrarily long-lived modes are not guaranteed for massive fields and there are examples when they do not exist \cite{Zinhailo:2024jzt,Konoplya:2005hr}. Moreover, the presence of a mass term alters the late-time signal decay: whereas massless fields exhibit power-law tails, massive fields produce oscillatory late-time tails \cite{Jing:2004zb,Koyama:2001qw,Moderski:2001tk,Rogatko:2007zz,Koyama:2001ee,Koyama:2000hj,Gibbons:2008gg,Gibbons:2008rs}. 
Furthermore, mass terms can effectively arise when a massless field propagates in a black hole immersed in an external magnetic field, as shown in \cite{Konoplya:2007yy,Konoplya:2008hj,Wu:2015fwa,Davlataliev:2024mjl}.

Summarizing, while perturbations of massless fields have been extensively studied recently in \cite{Konoplya:2025mvj,Momennia:2019cfd, Momennia:2019edt, Momennia:2018hsm, Konoplya:2020fwg, Malik:2024bmp}, there are no such studied for the massive fields, except the recent study of QNMs of a massive scalar field for the Schwarzschild-like Mannheim-Kazanas black holes in \cite{Becar:2023jtd}. No such analysis was done for the non-Schwarzschild black hole and wormholes solutions recently found in \cite{Jizba:2024owd}. The study of gray-body factors is absent for massive fields for all the three solutions in the Weyl gravity.  Given the significant differences introduced by massive terms in the perturbation equations, the study of QNMs and gray-body factors for massive fields in Weyl gravity is a natural extension of previous investigations. Understanding how the QNM spectrum evolves in this framework may provide crucial insights into the physical viability of Weyl gravity and its potential observational signatures in upcoming gravitational wave experiments.

The paper is organized as follows. In Sec. II, we provide a brief overview of spherically symmetric solutions in Weyl gravity, focusing on the Mannheim-Kazanas black hole and its generalization to wormhole geometries. In Sec. III, we derive the perturbation equation for a massive scalar field in these spacetimes and analyze the structure of the effective potential. In Sec. IV, we outline the semi-analytical JWKB method and time-domain integration technique, discussing their applicability to the QNM problem. In Sec. V, we compute the QNMs of a massive scalar field for both black hole and wormhole configurations, studying the impact of mass and the parameter $r_{0}$ on the spectrum. In Sec. VI, we analyze the grey-body factors and their connection to the QNM spectrum, verifying the validity of the correspondence between them. Finally, in Sec. VII, we summarize our findings and discuss potential extensions of this work.

\section{Mannheim-Kazanas and Jizba-Mudruňka solutions}\label{sec:MKsolution}

In Weyl conformal gravity, the effective cosmological constant naturally emerges as an integration constant within the vacuum field equations, without the need to introduce it explicitly \cite{Mannheim:1988dj}. This property distinguishes Weyl gravity from Einstein's general relativity, where the cosmological term must be inserted as an additional parameter. The fundamental action governing Weyl gravity is given by:
\begin{align}
S = \int d^4x \sqrt{-g} \, C^{abcd}C_{abcd},
\label{Weyl-action}
\end{align}
where \( g \) denotes the determinant of the metric tensor, and \( C_{abcd} \) represents the Weyl conformal tensor. Since the action is constructed solely from the square of the Weyl tensor, it remains invariant under local conformal transformations of the metric \( g_{\mu\nu}(x) \to \Omega^2(x) g_{\mu\nu}(x) \), making the theory fundamentally different from general relativity.

The field equations derived from this action take the form of fourth-order differential equations and are given by the vanishing of the Bach tensor:
\[
2 \partial_a \partial_d C^{ac}_{\phantom{ac}bc} + R_{ad} C^{ac}_{\phantom{ac}bc} = 0.
\]
Here, \( R_{ab} \) denotes the Ricci tensor, which is typically featured in second-order Einstein equations but enters Weyl gravity equations in a more complex manner. One remarkable feature of Weyl gravity is that Birkhoff's theorem still holds within this framework, ensuring that the only spherically symmetric vacuum solutions are static \cite{Riegert:1984zz}. The first black hole solution in this theory was derived by Mannheim and Kazanas in \cite{Mannheim:1988dj}, demonstrating the fundamental differences between Weyl gravity and Einstein's general relativity.

The Mannheim-Kazanas solution provides a Schwarzschild-like metric modified by additional terms that account for effective dark matter-like effects. Unlike standard Schwarzschild or Kerr solutions derived from Einstein's equations, the Mannheim-Kazanas metric arises from conformal gravity’s distinct field equations. This solution was originally proposed as an alternative explanation for galactic rotation curves without invoking the presence of dark matter \cite{Mannheim:1988dj}. Since Weyl gravity does not rely on an Einstein-Hilbert term, the field equations are purely conformal, leading to additional integration constants that naturally introduce cosmological and galactic-scale modifications.

Beyond the Mannheim-Kazanas solution, recent advancements in Weyl gravity have yielded new spherically symmetric solutions using the Newman-Penrose formalism \cite{Jizba:2024owd}. These solutions generalize the black hole metric by incorporating a parameter \( r_0 \) that modifies the radial function, leading to new classes of wormhole geometries. The line element for these solutions takes the form:

\begin{equation}
ds^2 = -f(r) dt^2 + \frac{dr^2}{f(r)} + (r^2+r_0^2)  \left( d\theta^2 + \sin^2\theta \, d\phi^2 \right),
\end{equation}
where the metric potential \( f(r) \)  has the following form:

\small
\begin{equation}\label{JMsolution}
f(r) =(r^2+r_0^2) \left(\frac{1 - \alpha + k r_0^2}{\rho^2(r)} - \frac{6 M+2\gamma r_0^2}{3\rho^3(r)} + \frac{\gamma}{\rho(r)} - k\right). 
\end{equation}
\normalsize
In this formulation, the function \( \rho(r) \) is defined as:
\begin{equation}\label{rhodef}
\frac{1}{\rho(r)}=\intop_r^{\infty}\frac{dr}{R^2(r)}=\frac{1}{r_0}\left(\frac{\pi}{2}-\arctan{\frac{r}{r_0}}\right),
\end{equation}
which asymptotically behaves as:
\[
\frac{\rho(r)}{r}=1+\Order{\frac{r_0}{r}}^2.
\]
Thus, in the asymptotic limit, the metric function simplifies to:
\begin{equation}
f(r)=1 - \alpha - \frac{2 M}{r} + \gamma r - k r^2 + \Order{\frac{r_0}{r}}^2,
\end{equation}
recovering the Mannheim-Kazanas solution in the limit \( r_0 \to 0 \).

In the approach of \cite{Konoplya:2025mvj}, the parameters \( M \) and \( \alpha \) are rewritten in terms of \( \beta \):
\begin{equation}\label{betadef}
\begin{array}{rcl}
    \alpha&=&3\beta\gamma + k r_0^2, \\
    M&=&\beta\left(1-\dfrac{3\beta\gamma}{2}\right) -\dfrac{\gamma r_0^2}{3}.
\end{array}
\end{equation}

This ensures consistency with the Mannheim-Kazanas parameters when taking the limit \( r_0 \to 0 \).

A crucial parameter in these solutions is \( r_0 \), which represents the minimum value of $r^2+r_0^2$, so that $r_{0}$ is the radius of the throat when $r_{0}$ is large enough to represent the wormhole. As \( r \to -\infty \), the wormhole asymptotics depend on the sign of \( f(r)/r^2 \). If an additional condition is satisfied,
\begin{equation}
\frac{M}{r_0} > \frac{1 - \alpha}{2\pi}-\gamma r_0\left(\frac{1}{3}-\frac{1}{2\pi^2}\right)+k r_0^2\frac{\pi^2-1}{2\pi^3},
\end{equation}
then the wormhole serves as a passage to an asymptotically de Sitter universe. Here, we focus on asymptotically de Sitter black holes and wormholes while disregarding the asymptotically anti-de Sitter solutions discussed in \cite{Jizba:2024owd,Konoplya:2025mvj}.

Since the expected values of the cosmological constant (associated with $k$) and the term explaining the rotation curves of galaxies ($\gamma$) are too small to produce a pronounced effect on the observed quasinormal spectrum, we analyze perturbations for all allowed values of these parameters including the regime of the near extreme black holes. This approach not only provides a theoretically complete description of the spectrum but also considers the possibility that the ratio between these constants may have been different in the early Universe.

\section{Wavelike equation}\label{sec:perturbations}

\begin{figure}
\resizebox{\linewidth}{!}{\includegraphics{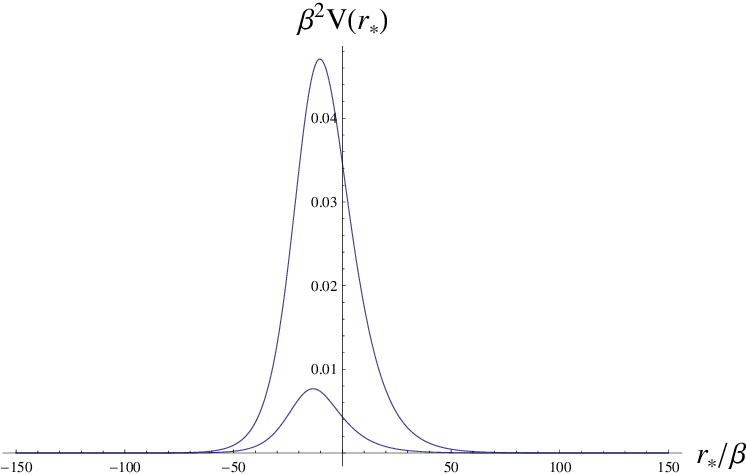}}
\caption{Effective potential as a function of the tortoise coordinate for scalar field perturbations: $\gamma =-0.3$, $\beta=1$, $\ell=1$, $r_{0}=2$, $k=0$, $\mu =0$ (lower) and $\mu =1$ (upper).}\label{fig:potentials}
\end{figure}

In this study, we investigate the behavior of a test massive scalar field in a curved spacetime, which not only provides insights into the dynamics of scalar perturbations but also serves as a useful model for understanding the qualitative features of massive fields with nonzero spin, such as massive fermions and gravitons. While it is well known that in the eikonal limit (\(\ell \to \infty\)), QNMs often exhibit a degree of universality, largely independent of the spin of the perturbing field, notable exceptions to this behavior have been identified \cite{Konoplya:2017wot}. These deviations frequently arise due to modifications induced by higher-order curvature corrections \cite{Bolokhov:2023dxq} or the influence of a nonzero cosmological constant \cite{Konoplya:2022gjp}. As a result, it remains an open question whether the presence of an effective mass term in gravitational perturbations in the Weyl gravity could introduce novel spectral characteristics, potentially revealing new physical effects in the quasinormal spectrum.

The evolution of a massive scalar field in a curved background is governed by the covariant Klein-Gordon equation:
\begin{equation}\label{coveqs}
\frac{1}{\sqrt{-g}}\partial_\mu \left(\sqrt{-g}g^{\mu \nu}\partial_\nu\Phi\right)=\mu^2 \Phi,
\end{equation}
where \(\mu\) represents the field's mass. Assuming a static, spherically symmetric background and employing a separation of variables, equation (\ref{coveqs}) can be reduced to a Schrödinger-like wave equation \cite{Kokkotas:1999bd,Berti:2009kk,Konoplya:2011qq}:
\begin{equation}\label{wave-equation}
\dfrac{d^2 \Psi}{dr_*^2}+(\omega^2-V(r))\Psi=0,
\end{equation}
where \(V(r)\) is the effective potential, and the tortoise coordinate \(r_*\) is introduced as:
\begin{equation}\label{tortoise}
dr_* \equiv \frac{dr}{f(r)}.
\end{equation}
Given the explicit form of the metric function in equation (\ref{JMsolution}), we observe that the ratio:
\begin{equation}\label{Pdef}
P(\rho(r))=\frac{f(r)}{r^2+r_0^2}, 
\end{equation}
depends solely on the function \(\rho(r)\). To further simplify our expressions, we define:
\begin{equation}\label{Prho}
P(\rho) = \frac{1 - 3\beta\gamma}{\rho^2} - \frac{2\beta-3\beta^2\gamma}{\rho^3} + \frac{\gamma}{\rho} - k.
\end{equation}

Importantly, this function remains independent of \(r_0\) once \(\alpha\) and \(M\) are parameterized in terms of \(\beta\) as described in equation (\ref{betadef}). Consequently, the tortoise coordinate transformation can also be rewritten in terms of \(\rho\):
\begin{equation}\label{tortoiseP}
dr_*=\frac{dr}{f(r)}=\frac{dr}{(r^2+r_0^2) P(\rho(r))}=\frac{d\rho}{\rho^2P(\rho)}.
\end{equation}
The effective potential governing the scalar (\(s=0\)) perturbations in this background takes the form:
\begin{eqnarray}\label{potentialScalar}
V(r) &=& (\ell(\ell+1) +\mu^2 (r^2+r_0^2)) P(\rho(r)) \nonumber \\
&+& \frac{1}{(r^2+r_0^2)}\frac{d^2 \sqrt{r^2+r_0^2}} {dr_*^2}.
\end{eqnarray}
where \(\ell = 0,1,2,\dots\) represents the multipole number, which appears naturally from the separation of angular variables.

The characteristics of the effective potential for different values of \(\mu\) and \(\ell\) are illustrated in Figure \ref{fig:potentials}, demonstrating how the mass term alters the structure of the potential well and its asymptotic behavior. The analysis of these modifications is crucial for assessing the stability of the corresponding solutions and for understanding potential observational imprints of massive QNMs in Weyl gravity.

\section{JWKB method and time-domain integration }\label{sec:numerical}
The QNMs of asymptotically de Sitter black holes correspond to the characteristic oscillation frequencies, $\omega$, of the wave-like equation (\ref{wave-equation}), subject to the following boundary conditions:
\begin{equation}\label{boundaryconditions}
\Psi \propto e^{+ \imo \omega r_*}, \quad  r_* \to  +\infty \quad \text{(de Sitter horizon)},
\end{equation}
\begin{equation}\label{boundaryconditions2}
\Psi \propto e^{- \imo \omega r_*}, \quad  r_* \to - \infty  \quad (event \quad  horizon),
\end{equation}
which ensure a purely ingoing wave at the event horizon and a purely outgoing wave at the de Sitter horizon. 

These boundary conditions are also applicable to traversable wormholes, where, in the case of an asymptotically flat external region, the coordinate limit $r_*\to-\infty$ corresponds to spatial infinity in the second asymptotic universe \cite{Konoplya:2005et}. For wormholes with de Sitter asymptotics on one or both sides, purely outgoing conditions at the respective de Sitter horizons are imposed in a similar manner. 

Among the various techniques available for calculating quasinormal frequencies, the JWKB approximation stands out as one of the most commonly used methods, owing to its efficiency, automation, and generally sufficient accuracy in many cases (see, e.g., \cite{Barrau:2019swg,Stashko:2024wuq,Konoplya:2001ji,Zinhailo:2019rwd,Xiong:2021cth,Skvortsova:2024atk,Konoplya:2006ar,Skvortsova:2024wly,Malik:2024nhy,Hamil:2024nrv,Liu:2024wch,Kodama:2009bf}). The JWKB method relies on approximating the effective potential near its peak using a Taylor expansion and matching asymptotic solutions that satisfy the QNM boundary conditions (\ref{boundaryconditions}) through two classical turning points. As a result, this method is particularly effective when the effective potential forms a single-barrier potential. 

Fortunately, in the scenario where the field mass is not negligible and the parameters $k$ or $\gamma$ are nonzero, the effective potential exhibits a barrier-like structure, making the JWKB approach a reliable tool for obtaining quasinormal frequencies with good accuracy \cite{Fontana:2020syy,Dubinsky:2024hmn}. The general JWKB formula expresses quasinormal frequencies as an expansion around the eikonal limit \cite{Konoplya:2019hlu}:
\begin{eqnarray}\label{WKBformula-spherical}
\omega^2 &=& V_0 + A_2(\K^2) + A_4(\K^2) + \ldots \\ \nonumber
&-& \imo \K \sqrt{-2V_2} \left( 1 + A_3(\K^2) + A_5(\K^2) + \ldots \right),
\end{eqnarray}
where the quantization condition for quasinormal frequencies imposes
\[
\K = n + \frac{1}{2}, \quad n = 0, 1, 2, \ldots,
\]
with $n$ denoting the overtone number. Here, $V_0$ represents the peak value of the effective potential, while $V_2$ corresponds to its second derivative at this point. The correction terms $A_i$, with $i = 2, 3, 4, \dots$, encode higher-order JWKB corrections beyond the eikonal approximation, incorporating derivatives of the potential up to order $2i$. Explicit expressions for these corrections were derived in \cite{Iyer:1986np} for the second and third JWKB orders, in \cite{Konoplya:2003ii} for the 4th-6th orders, and in \cite{Matyjasek:2017psv} for the 7th-13th orders.

While the JWKB method provides an analytical approximation for quasinormal frequencies, numerical time-domain approaches offer an alternative means of extracting the dominant frequencies from black hole perturbations. One of the most widely adopted numerical techniques in this context is the Gundlach-Price-Pullin method \cite{Gundlach:1993tp}, which enables the computation of the full time evolution of perturbations. This approach has been extensively applied to black hole spacetimes (see, for example, \cite{Churilova:2021tgn,Konoplya:2013sba,Momennia:2022tug,Varghese:2011ku,Skvortsova:2024atk,Aneesh:2018hlp,Malik:2024nhy}) and allows for a precise extraction of quasinormal frequencies. 

The method is based on discretizing the wave equation, which takes the general form:
\[
\frac{\partial^2 \Psi}{\partial t^2} - \frac{\partial^2 \Psi}{\partial r_*^2} + V(r_*)\Psi = 0,
\]
where $\Psi$ is the perturbation field, $r_*$ is the tortoise coordinate, and $V(r_*)$ is the effective potential. By introducing characteristic coordinates $u = t - r_*$ and $v = t + r_*$, the wave equation can be discretized on a numerical grid, leading to the finite-difference equation:
\begin{eqnarray}
\Psi\left(N\right)&=&\Psi\left(W\right)+\Psi\left(E\right)-\Psi\left(S\right)
\nonumber\\&&
-\Delta^2V\left(S\right)\frac{\Psi\left(W\right)+\Psi\left(E\right)}{8}+{\cal O}\left(\Delta^4\right),\label{Discretization}
\end{eqnarray}
where the grid points are defined as follows:
\[
N\equiv\left(u+\Delta,v+\Delta\right), \quad W\equiv\left(u+\Delta,v\right),
\]
\[
E\equiv\left(u,v+\Delta\right), \quad S\equiv\left(u,v\right).
\]
Once the time-domain profiles of the perturbations are obtained, the Prony method is applied to extract quasinormal frequencies from the late-time signal. This technique involves fitting the numerical waveform to a sum of exponentially damped sinusoids, allowing for an accurate determination of the dominant modes.

Overall, the combination of JWKB approximations and time-domain numerical methods provides a comprehensive approach to the study of QNMs in black hole spacetimes. While JWKB methods are efficient and analytically tractable, numerical simulations are indispensable for capturing the full complexity of wave dynamics.

\section{Quasinormal Modes}

\begin{widetext}
\begin{table*}
    \centering
    \renewcommand{\arraystretch}{1.2} 
    \setlength{\tabcolsep}{10pt} 
    \begin{tabular}{|c|c|c|c|c|}
        \hline
        \hline
        $\mu$ & $\gamma=-0.1$ & $\gamma=-0.2$ & $\gamma=-0.3$ & $\gamma=-0.32$ \\ 
        \hline
        0.0  & 0.075727 - 0.104860 i & 0.042975 - 0.092209 i & 0.017385 - 0.058034 i & 0.021757 - 0.029204 i \\
        0.2  & 0.126230 - 0.057254 i & 0.065015 - 0.053199 i & 0.023167 - 0.028687 i & 0.014599 - 0.018073 i \\
        0.4  & 0.234752 - 0.040269 i & 0.154137 - 0.047422 i & 0.068915 - 0.028955 i & 0.042635 - 0.018894 i \\
        0.6  & 0.349024 - 0.040829 i & 0.234936 - 0.046366 i & 0.107880 - 0.028773 i & 0.067212 - 0.018835 i \\
        0.8  & 0.466131 - 0.041499 i & 0.314955 - 0.046320 i & 0.145919 - 0.028791 i & 0.091103 - 0.018837 i \\
        1.0  & 0.582690 - 0.041651 i & 0.394824 - 0.046428 i & 0.183615 - 0.028806 i & 0.114736 - 0.018842 i \\
        2.0  & 1.167006 - 0.042042 i & 0.793150 - 0.046682 i & 0.370507 - 0.028842 i & 0.231755 - 0.018851 i \\
        6.0  & 3.502650 - 0.042139 i & 2.382855 - 0.046766 i & 1.114471 - 0.028857 i & 0.697296 - 0.018855 i \\
       10.0  & 5.837971 - 0.042147 i & 3.971886 - 0.046772 i & 1.857847 - 0.028858 i & 1.162431 - 0.018855 i \\
       14.0  & 8.173246 - 0.042149 i & 5.560819 - 0.046774 i & 2.601137 - 0.028858 i & 1.627507 - 0.018855 i \\
       18.0  & 10.508504 - 0.042150 i & 7.149719 - 0.046775 i & 3.344400 - 0.028858 i & 2.092565 - 0.018855 i \\
       22.0  & 12.843755 - 0.042150 i & 8.738604 - 0.046775 i & 4.087650 - 0.028858 i & 2.557613 - 0.018855 i \\
        \hline
        \hline
    \end{tabular}
    \caption{Fundamental QNMs of the Schwarzschild branch for different values of $\mu$ and $\gamma$: $\ell=0$, $k=0$, $r_{0}=0$, $\beta =1$.}
    \label{tab1}
\end{table*}
\begin{table*}
    \centering
    \renewcommand{\arraystretch}{1.2} 
    \setlength{\tabcolsep}{10pt} 
    \begin{tabular}{|c|c|c|c|c|}
        \hline
         \hline
        $\mu$ & $\gamma=-0.1$ & $\gamma=-0.2$ & $\gamma=-0.3$ & $\gamma=-0.32$ \\ 
        \hline
        0.0  & 0.235087 - 0.084329 i & 0.169972 - 0.064213 i & 0.081406 - 0.031052 i & 0.051132 - 0.019411 i \\
        0.2  & 0.252409 - 0.076361 i & 0.184780 - 0.060013 i & 0.089343 - 0.030466 i & 0.056142 - 0.019261 i \\
        0.4  & 0.309497 - 0.057414 i & 0.226865 - 0.053011 i & 0.109857 - 0.029626 i & 0.069039 - 0.019049 i \\
        0.6  & 0.398933 - 0.046463 i & 0.286080 - 0.049128 i & 0.137542 - 0.029168 i & 0.086377 - 0.018934 i \\
        0.8  & 0.501470 - 0.043047 i & 0.353768 - 0.047555 i & 0.168933 - 0.028978 i & 0.106011 - 0.018886 i \\
        1.0  & 0.610198 - 0.042232 i & 0.425910 - 0.047015 i & 0.202322 - 0.028902 i & 0.126881 - 0.018867 i \\
        2.0  & 1.179936 - 0.042057 i & 0.808522 - 0.046717 i & 0.380057 - 0.028850 i & 0.237985 - 0.018853 i \\
        6.0  & 3.506889 - 0.042140 i & 2.387951 - 0.046766 i & 1.117673 - 0.028857 i & 0.699389 - 0.018855 i \\
       10.0  & 5.840511 - 0.042147 i & 3.974943 - 0.046772 i & 1.859768 - 0.028858 i & 1.163687 - 0.018855 i \\
       14.0  & 8.175059 - 0.042149 i & 5.563002 - 0.046774 i & 2.602510 - 0.028858 i & 1.628405 - 0.018855 i \\
       18.0  & 10.509914 - 0.042150 i & 7.151417 - 0.046775 i & 3.345468 - 0.028858 i & 2.093263 - 0.018855 i \\
       22.0  & 12.844909 - 0.042150 i & 8.739993 - 0.046775 i & 4.088524 - 0.028858 i & 2.558184 - 0.018855 i \\
        \hline
         \hline
    \end{tabular}
    \caption{Fundamental QNMs of the Schwarzschild branch for different values of $\gamma$ and $\mu$: $\ell=1$, $k=0$, $r_{0}=0$, $\beta =1$.}
    \label{tab2}
\end{table*}
\begin{table*}
    \centering
    \renewcommand{\arraystretch}{1.2} 
    \setlength{\tabcolsep}{10pt} 
    \begin{tabular}{|c|c|c|c|}
        \hline
        \hline
        $\mu$ & $\gamma=-0.1$, $k=0.01$ & $\gamma=-0.1$, $k=0.02$ & $\gamma=-0.1$, $k=0.025$ \\ 
        \hline
        0.0  & 0.051076 - 0.092968 i & 0.022930 - 0.065779 i & 0.014913 - 0.016162 i \\
        0.2  & 0.072192 - 0.058755 i & 0.030466 - 0.039611 i & 0.011026 - 0.014023 i \\
        0.4  & 0.161345 - 0.050453 i & 0.089044 - 0.036180 i & 0.033506 - 0.015089 i \\
        0.6  & 0.245363 - 0.049219 i & 0.138478 - 0.035824 i & 0.052935 - 0.015060 i \\
        0.8  & 0.328730 - 0.049164 i & 0.187000 - 0.035820 i & 0.071793 - 0.015060 i \\
        1.0  & 0.411966 - 0.049272 i & 0.235124 - 0.035841 i & 0.090439 - 0.015061 i \\
        2.0  & 0.827325 - 0.049539 i & 0.474024 - 0.035903 i & 0.182729 - 0.015065 i \\
        6.0  & 2.485330 - 0.049627 i & 1.425519 - 0.035928 i & 0.549829 - 0.015067 i \\
       10.0  & 4.142672 - 0.049634 i & 2.376328 - 0.035930 i & 0.916600 - 0.015067 i \\
       14.0  & 5.799917 - 0.049636 i & 3.327038 - 0.035931 i & 1.283325 - 0.015067 i \\
       18.0  & 7.457129 - 0.049637 i & 4.277714 - 0.035931 i & 1.650034 - 0.015067 i \\
       22.0  & 9.114326 - 0.049637 i & 5.228376 - 0.035931 i & 2.016736 - 0.015067 i \\
        \hline
        \hline
    \end{tabular}
    \caption{Fundamental QNMs of the Schwarzschild branch for different values of $\gamma$, $\mu$ and $k$: $\ell=0$, $r_{0}=0$, $\beta=1$.}
    \label{tab3}
\end{table*}
\begin{table*}
    \centering
    \renewcommand{\arraystretch}{1.2} 
    \setlength{\tabcolsep}{10pt} 
    \begin{tabular}{|c|c|c|c|}
        \hline
        \hline
        $\mu$ & $\gamma=-0.1$, $k=0.01$ & $\gamma=-0.1$, $k=0.02$ & $\gamma=-0.1$, $k=0.025$ \\ 
        \hline
        0.0  & 0.177744 - 0.067134 i & 0.103975 - 0.039931 i & 0.040377 - 0.015315 i \\
        0.2  & 0.192926 - 0.062697 i & 0.113974 - 0.038787 i & 0.044329 - 0.015244 i \\
        0.4  & 0.236515 - 0.055368 i & 0.139987 - 0.037195 i & 0.054501 - 0.015145 i \\
        0.6  & 0.298093 - 0.051431 i & 0.175295 - 0.036346 i & 0.068172 - 0.015093 i \\
        0.8  & 0.368588 - 0.049954 i & 0.215437 - 0.036021 i & 0.083650 - 0.015073 i \\
        1.0  & 0.443837 - 0.049510 i & 0.258176 - 0.035910 i & 0.100103 - 0.015066 i \\
        2.0  & 0.843020 - 0.049469 i & 0.485721 - 0.035884 i & 0.187691 - 0.015064 i \\
        6.0  & 2.490526 - 0.049615 i & 1.429431 - 0.035925 i & 0.551496 - 0.015067 i \\
       10.0  & 4.145788 - 0.049629 i & 2.378676 - 0.035929 i & 0.917602 - 0.015067 i \\
       14.0  & 5.802142 - 0.049634 i & 3.328715 - 0.035930 i & 1.284040 - 0.015067 i \\
       18.0  & 7.458860 - 0.049635 i & 4.279019 - 0.035931 i & 1.650590 - 0.015067 i \\
       22.0  & 9.115742 - 0.049636 i & 5.229443 - 0.035931 i & 2.017191 - 0.015067 i \\
        \hline
        \hline
    \end{tabular}
    \caption{Fundamental QNMs of the Schwarzschild branch for different values of $\gamma$, $\mu$ and $k$: $\ell=1$, $r_{0}=0$, $\beta=1$.}
    \label{tab4}
\end{table*}
\begin{table*}
    \centering
    \renewcommand{\arraystretch}{1.2} 
    \setlength{\tabcolsep}{10pt} 
    \begin{tabular}{|c|c|c|c|c|}
        \hline
        \hline
        $\mu$  &  $\gamma=-0.1$ & $\gamma=-0.2$ & $\gamma=-0.25$ & $\gamma=-0.32$ \\ 
        \hline
        0.0  & 0.229725 - 0.086104 i & 0.167374 - 0.064521 i & 0.130047 - 0.050336 i & 0.051045 - 0.019406 i \\
        0.2  & 0.257311 - 0.078426 i & 0.189671 - 0.060660 i & 0.148044 - 0.048292 i & 0.058234 - 0.019277 i \\
        0.4  & 0.332585 - 0.062833 i & 0.246279 - 0.054959 i & 0.192512 - 0.045531 i & 0.075794 - 0.019106 i \\
        0.6  & 0.435100 - 0.052743 i & 0.319693 - 0.051548 i & 0.249668 - 0.043926 i & 0.098303 - 0.019006 i \\
        0.8  & 0.548610 - 0.047873 i & 0.400557 - 0.049774 i & 0.312578 - 0.043090 i & 0.123078 - 0.018955 i \\
        1.0  & 0.667030 - 0.045456 i & 0.485120 - 0.048813 i & 0.378381 - 0.042630 i & 0.148994 - 0.018926 i \\
        2.0  & 1.284042 - 0.042641 i & 0.927205 - 0.047432 i & 0.722539 - 0.041946 i & 0.284528 - 0.018883 i \\
        6.0  & 3.809008 - 0.042115 i & 2.742481 - 0.047032 i & 2.136298 - 0.041734 i & 0.841245 - 0.018868 i \\
       10.0  & 6.342701 - 0.042083 i & 4.565564 - 0.047001 i & 3.556292 - 0.041717 i & 1.400417 - 0.018867 i \\
       14.0  & 8.877609 - 0.042074 i & 6.389768 - 0.046993 i & 4.977186 - 0.041712 i & 1.959943 - 0.018867 i \\
       18.0  & 11.412919 - 0.042071 i & 8.214347 - 0.046989 i & 6.398381 - 0.041710 i & 2.519588 - 0.018867 i \\
       22.0  & 13.948412 - 0.042069 i & 10.039096 - 0.046987 i & 7.819712 - 0.041709 i & 3.079287 - 0.018867 i \\
     \hline
     \hline
    \end{tabular}
    \caption{Fundamental QNMs of the Schwarzschild branch for different values of $\gamma$ and $\mu$: $\ell=1$, $r_{0}=3.14$, $k=0$, $\beta=1$.}
    \label{tab5}
\end{table*}
\begin{table*}
    \centering
    \renewcommand{\arraystretch}{1.2} 
    \setlength{\tabcolsep}{10pt} 
    \begin{tabular}{|c|c|c|c|c|}
        \hline
        \hline
        $\mu$  &  $\gamma=-0.1$ & $\gamma=-0.2$ & $\gamma=-0.25$ & $\gamma=-0.32$ \\ 
        \hline
        0.0  & 0.112299 - 0.140957 i & 0.069409 - 0.109634 i & 0.047421 - 0.082373 i & 0.003335 - 0.007090 i \\
        0.2  & 0.152663 - 0.077038 i & 0.117171 - 0.069443 i & 0.087464 - 0.052773 i & 0.033904 - 0.019446 i \\
        0.4  & 0.332005 - 0.067850 i & 0.246371 - 0.056664 i & 0.192120 - 0.046390 i & 0.075220 - 0.019166 i \\
        0.6  & 0.493490 - 0.063722 i & 0.369351 - 0.054116 i & 0.289888 - 0.044936 i & 0.114654 - 0.019057 i \\
        0.8  & 0.655219 - 0.062272 i & 0.491806 - 0.053065 i & 0.386975 - 0.044345 i & 0.153686 - 0.019012 i \\
        1.0  & 0.817329 - 0.061667 i & 0.614224 - 0.052546 i & 0.483889 - 0.044049 i & 0.192568 - 0.018991 i \\
        2.0  & 1.630352 - 0.061110 i & 1.226664 - 0.051811 i & 0.968003 - 0.043624 i & 0.386346 - 0.018960 i \\
        6.0  & 4.887825 - 0.061059 i & 3.678147 - 0.051584 i & 2.904037 - 0.043490 i & 1.160101 - 0.018951 i \\
       10.0  & 8.145982 - 0.061059 i & 6.129990 - 0.051566 i & 4.840059 - 0.043479 i & 1.933643 - 0.018950 i \\
       14.0  & 11.404225 - 0.061059 i & 8.581888 - 0.051561 i & 6.776082 - 0.043476 i & 2.707154 - 0.018950 i \\
       18.0  & 14.662496 - 0.061059 i & 11.033803 - 0.051559 i & 8.712104 - 0.043474 i & 3.480656 - 0.018950 i \\
       22.0  & 17.920779 - 0.061059 i & 13.485727 - 0.051558 i & 10.648127 - 0.043474 i & 4.254153 - 0.018950 i \\
        \hline
        \hline
    \end{tabular}
    \caption{Fundamental QNMs of the Schwarzschild branch for different values of $\gamma$ and $\mu$: $\ell=0$, $r_{0}=5$, $k=0$, $\beta=1$.}
    \label{tab6}
\end{table*}
\begin{table*}
    \centering
    \renewcommand{\arraystretch}{1.2} 
    \setlength{\tabcolsep}{10pt} 
    \begin{tabular}{|c|c|c|c|c|}
        \hline
        \hline
       $\mu$  &  $\gamma=-0.1$ & $\gamma=-0.2$ & $\gamma=-0.25$ & $\gamma=-0.32$ \\ 
        \hline
        0.0  & 0.241004 - 0.098608 i  & 0.170751 - 0.071353 i  & 0.130798 - 0.053524 i  & 0.051101 - 0.019542 i \\
        0.2  & 0.289548 - 0.085939 i  & 0.210912 - 0.064515 i  & 0.163874 - 0.050156 i  & 0.064175 - 0.019j387 i \\
        0.4  & 0.404212 - 0.072797 i  & 0.300528 - 0.058148 i  & 0.235355 - 0.046965 i  & 0.092842 - 0.019190 i \\
        0.6  & 0.544112 - 0.066733 i  & 0.407305 - 0.055245 i  & 0.320107 - 0.045464 i  & 0.126909 - 0.019088 i \\
        0.8  & 0.693921 - 0.063942 i  & 0.520891 - 0.053833 i  & 0.410100 - 0.044722 i  & 0.163033 - 0.019036 i \\
        1.0  & 0.848487 - 0.062628 i  & 0.637743 - 0.053078 i  & 0.502578 - 0.044319 i  & 0.200108 - 0.019008 i \\
        2.0  & 1.645901 - 0.061226 i  & 1.238598 - 0.051955 i  & 0.977486 - 0.043703 i  & 0.390160 - 0.018965 i \\
        6.0  & 4.892976 - 0.061064 i  & 3.682142 - 0.051601 i  & 2.907213 - 0.043499 i  & 1.161377 - 0.018951 i \\
       10.0  & 8.149071 - 0.061060 i  & 6.132388 - 0.051572 i  & 4.841965 - 0.043482 i  & 1.934409 - 0.018950 i \\
       14.0  & 11.406430 - 0.061059 i & 8.583600 - 0.051564 i  & 6.777443 - 0.043477 i  & 2.707701 - 0.018950 i \\
       18.0  & 14.664211 - 0.061059 i & 11.035135 - 0.051561 i & 8.713163 - 0.043475 i  & 3.481081 - 0.018950 i \\
       22.0  & 17.922183 - 0.061059 i & 13.486817 - 0.051559 i & 10.648993 - 0.043474 i & 4.254501 - 0.018950 i \\
        \hline
        \hline
    \end{tabular}
    \caption{Fundamental QNMs of the Schwarzschild branch for different values of $\gamma$ and $\mu$: $\ell=1$, $r_{0}=5$, $k=0$, $\beta=1$.}
    \label{tab7}
\end{table*}
\begin{table*}
    \centering
    \renewcommand{\arraystretch}{1.2} 
    \setlength{\tabcolsep}{10pt} 
    \begin{tabular}{|c|c|c|c|c|}
        \hline
        \hline
       $\mu$ & $\gamma=-0.25$, $\ell=1$ & $\gamma=-0.32$, $\ell=1$ & $\gamma=-0.25$, $\ell=2$ & $\gamma=-0.32$, $\ell=2$ \\ 
        \hline
        0.0  & 0.144593 - 0.059099 i & 0.051626 - 0.020193 i & 0.241194 - 0.052179 i & 0.092892 - 0.019575 i \\  
        0.2  & 0.203702 - 0.050260 i & 0.076360 - 0.019440 i & 0.279853 - 0.048933 i & 0.108441 - 0.019342 i \\  
        0.4  & 0.329601 - 0.047642 i & 0.123840 - 0.019212 i & 0.378488 - 0.046308 i & 0.145675 - 0.019162 i \\  
        0.6  & 0.472064 - 0.048110 i & 0.176655 - 0.019179 i & 0.505714 - 0.046626 i & 0.192512 - 0.019129 i \\  
        0.8  & 0.619508 - 0.048611 i & 0.231164 - 0.019174 i & 0.644719 - 0.047441 i & 0.243460 - 0.019133 i \\  
        1.0  & 0.768868 - 0.048928 i & 0.286396 - 0.019173 i & 0.788953 - 0.048062 i & 0.296391 - 0.019142 i \\  
        2.0  & 1.523562 - 0.049442 i & 0.565913 - 0.019175 i & 1.533515 - 0.049186 i & 0.571022 - 0.019165 i \\  
        6.0  & 4.558480 - 0.049614 i & 1.691593 - 0.019176 i & 4.561787 - 0.049584 i & 1.693307 - 0.019175 i \\  
       10.0  & 7.595853 - 0.049628 i & 2.818501 - 0.019176 i & 7.597836 - 0.049617 i & 2.819530 - 0.019176 i \\  
       14.0  & 10.633572 - 0.049631 i & 3.945585 - 0.019176 i & 10.634988 - 0.049626 i & 3.946320 - 0.019176 i \\  
       18.0  & 13.671406 - 0.049633 i & 5.072727 - 0.019176 i & 13.672508 - 0.049630 i & 5.073299 - 0.019176 i \\  
       22.0  & 16.709293 - 0.049634 i & 6.199896 - 0.019176 i & 16.710194 - 0.049632 i & 6.200364 - 0.019176 i \\  
        \hline
        \hline
    \end{tabular}
    \caption{QNMs for different values of $\gamma$, $\ell$ and $\mu$: $r_{0}=6.28$, $k=0$, $\beta =1$. }
    \label{tab:QNM_data_final}
\end{table*}
\end{widetext}

\begin{figure}
\resizebox{\linewidth}{!}{\includegraphics{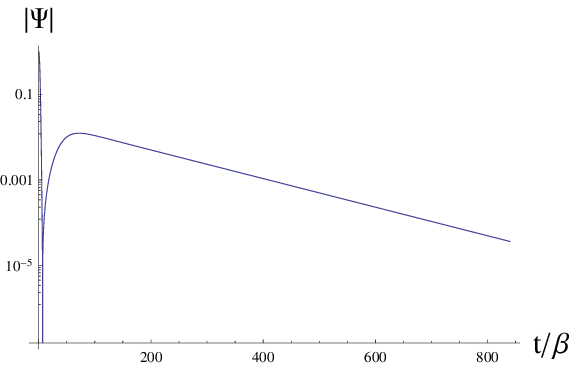}}
\caption{Semi-logarithmic time-domain profile for
$r_{0}=2$, $\gamma=-0.3$, $\mu =0.1$, $\ell=0$, $\beta=1$. Prony method gives $\omega_{0} = - 0.00770512 i$ for the fundamental mode of the de Sitter branch.}\label{fig:TD0}
\end{figure}
\begin{figure*}
\resizebox{\linewidth}{!}{\includegraphics{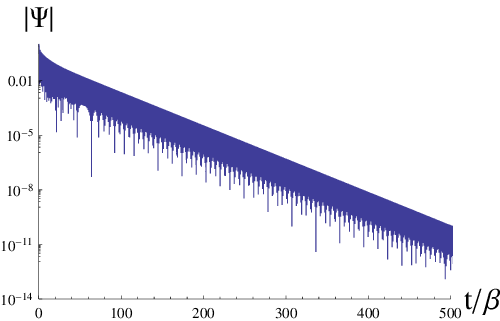}~~~\includegraphics{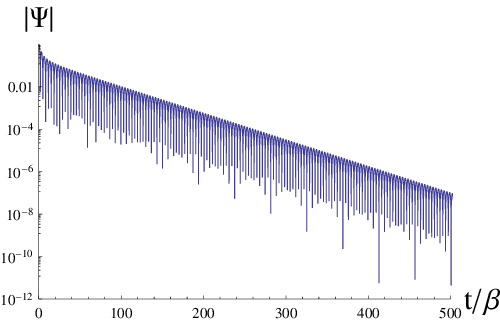}}
\caption{Semi-logarithmic time-domain profiles. \textit{Left plot:} $r_{0}=0$, $\gamma=-0.1$, $\mu =5$, $\ell=0$. The 6th order JWKB data gives:  $\omega_{0} = 2.918799 - 0.042134 i$ and $\omega_{1}= 2.918152 - 0.126393 i$, while the time-domain integration with the Prony method leads to the very close frequencies $\omega_{0} =2.9189 - 0.0421296 i$ and $\omega_{1}= 2.91801 - 0.126426 i$. \textit{Right plot:} 
$r_{0}=0$, $\gamma=-0.3$, $\mu =5$, $\ell=1$, we have $\omega_{0} = 0.932434 - 0.028856 i$ and  $\omega_{1} = 0.932386 -  0.086854 i$ from the time-domain integration and the JWKB method gives $\omega_{0} = 0.932431 - 0.028856 i$, $\omega_{1} = 0.932382 - 0.086568 i$. We can see that the two dominant frequencies have concurrent real oscillation frequencies.}\label{fig:TD1}
\end{figure*}
\begin{figure*}
\resizebox{\linewidth}{!}{\includegraphics{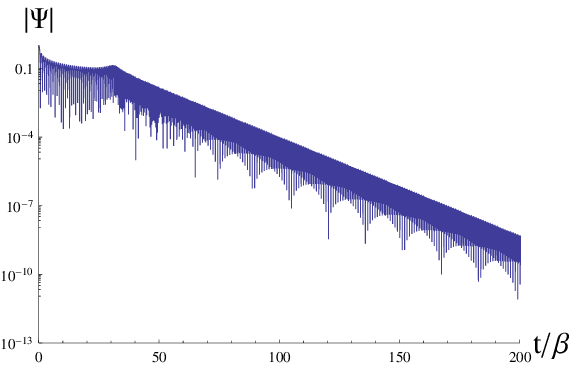}~~~\includegraphics{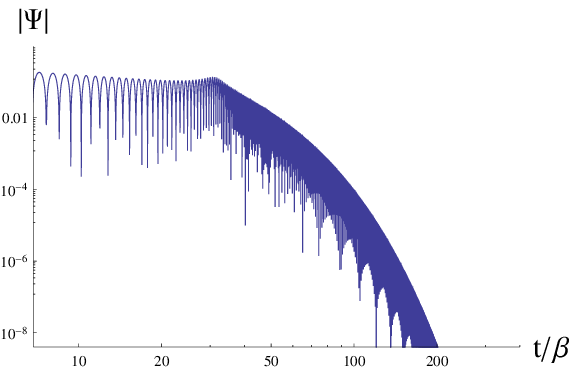}}
\caption{Semi-logarithmic (left) and logarithmic time-domain profiles. \textit{Left plot:} $r_{0}=6.28$, $\gamma=-0.1$, $\mu =5$, $\ell=1$. The 6th order JWKB data gives:  $\omega_{0} = 9.417267-0.099480 i$ and $\omega_{1}= 9.416725-0.298460 i$, while the time-domain integration with the Prony method leads to the very close frequencies $\omega_{0} = 9.42075 - 0.099366 i$ and $\omega_{1}= 9.41822 - 0.299113 i$.}\label{fig:TD2}
\end{figure*}

The QNMs of asymptotically de Sitter black holes have been computed using various numerical and semi-analytical methods \cite{Konoplya:2004uk, Zhidenko:2003wq, Dubinsky:2024gwo, Cardoso:2004up, Malik:2024cgb, Hatsuda:2020sbn, Cuyubamba:2016cug, Konoplya:2007zx, Dubinsky:2024hmn, Konoplya:2013sba}. The most notable effect of a nonzero cosmological constant is the emergence of a new branch of modes \cite{Konoplya:2022xid}, which corresponds to the quasinormal spectrum of empty de Sitter spacetime \cite{Lopez-Ortega:2006aal,Lopez-Ortega:2007vlo,Lopez-Ortega:2012xvr}, modified by the presence of a black hole. The exact analytical expression for the quasinormal frequencies of a massive scalar field in the pure de Sitter spacetime was derived in \cite{Lopez-Ortega:2012xvr}.

Thus, in the limit \( M \to 0 \), the quasinormal frequencies of the de Sitter spacetime take the form \cite{Lopez-Ortega:2012xvr}:
\begin{equation}
    i\omega_n k^{-1/2} = \ell + 2n + \frac{3}{2} \pm \sqrt{\frac{9}{4} - \mu^2 k^{-1}},
\end{equation}
for
\begin{equation}
    \frac{9}{4} > \mu^2 k^{-1},
\end{equation}
and
\begin{equation}
    i\omega_n k^{-1/2} = \ell + 2n + \frac{3}{2} \pm i \sqrt{\mu^2 k^{-1} - \frac{9}{4}},
\end{equation}
for
\begin{equation}
    \frac{9}{4} < \mu^2 k^{-1}.
\end{equation}
For small values of \( \mu \), the quasinormal frequencies of pure de Sitter spacetime remain purely imaginary, consistent with the massless case \cite{Konoplya:2022xid}, which is illustrated here on fig. \ref{fig:TD0}. When \( \gamma=0 \) and the black hole mass approaches zero, the QNMs of the black hole smoothly transition to those of pure de Sitter spacetime.

Quasinormal modes of the Schwarzschild branch were found with the WKB method with Padé approximants and are shown in tables I-VIII.
The parameter \( \gamma \) plays a role similar to that of a cosmological constant, inducing the formation of a de Sitter-like horizon in the far region even when \( k=0 \). This suggests that, for sufficiently small cosmological constant, the primary focus should be on studying the dependence of quasinormal frequencies on the parameter \( \gamma \) and the mass of the field \( \mu \). Since the de Sitter and \( \gamma \)-branch modes cannot be accurately determined using the JWKB method, we begin by analyzing the Schwarzschild branch of modes. These are computed with high precision using the sixth-order JWKB method with Padé approximants.

In the case \( r_0 = 0 \), the solution reduces to the Mannheim-Kazanas metric, for which QNMs of a massive scalar field have been recently investigated in \cite{Becar:2023jtd}. In Tables 1–4, we confirm their findings and further demonstrate that increasing \( \mu \) decreases the damping rate. However, contrary to the case of asymptotically flat black holes \cite{Konoplya:2004wg}, the damping rate does not vanish completely. Instead, it asymptotically reaches a minimal limiting value for large field masses, while the real oscillation frequency scales proportionally with \( \mu \).
Notice that similar absence of arbitrary long lived modes takes place in brane-wrold models, as was shown in \cite{Zinhailo:2024jzt}.

Following \cite{Konoplya:2004wg}, we can demonstrate that arbitrarily long-lived modes (quasi-resonances) are forbidden in our case. By multiplying the wave equation by $\Psi^*$ and integrating over the spatial domain, we obtain:
\begin{widetext}
\begin{equation}
\int_{-r_1}^{+r_2} \Psi^* \left( \Psi'' + (\omega^2 - V) \Psi \right) dr^* = \omega^2 \int_{-r_1}^{+r_2} |\Psi|^2 dr^* 
+ \int_{-r_1}^{+r_2} (|\Psi'|^2 + V |\Psi|^2) dr^* + \Psi^* \Psi' \bigg|_{r^* = -r_1}^{r^* = +r_2} = 0.
\label{eq:13}
\end{equation}
The boundary conditions are set at the event horizon ($r_1$) and the de Sitter-like horizon ($r_2$), induced by the nonzero effective dark matter term $\gamma < 0$. Under these conditions, we obtain:
\begin{equation}
\Psi^* \Psi' \bigg|^{r^* = +r_2}_{r^* = -r_1} = i \exp(-2\text{Im}(\chi)r_2) r_2^{2M \text{Im}(\chi)\mu^2 / |\chi|^2}
\left( \chi |C_+|^2 + \mathcal{O} \left( \frac{1}{r_2} \right) r_2 \right)
+ i \exp(-2\text{Im}(\omega) r_1) \left( \omega |C_-|^2 + \mathcal{O} \left( \frac{1}{r_1} \right) r_1 \right),
\label{eq:14}
\end{equation}
\end{widetext}
where $\chi = \sqrt{\omega^2 -\mu^2}$. When the field mass $\mu$ is nonzero, frequencies with $\text{Im}(\omega) = 0$ are not necessarily excluded, as equation (\ref{eq:13}) is satisfied under the condition:
\begin{equation}
\text{Re}(\chi) |C_+|^2 = 0, \quad \text{and} \quad \omega |C_-|^2 = 0.
\label{eq:15}
\end{equation}
This implies that no wave falls into the event horizon ($C_- = 0$), and consequently, there is no energy transmission to infinity, otherwise we would have $\text{Re}(\chi) = 0$. On the other hand, equation (\ref{eq:15}) suggests that quasi-resonances can exist if the potential does not vanish at spatial infinity or de Sitter horizon. However, for compact objects in Weyl gravity, the spacetime asymptotics always exhibit a de Sitter-like behavior (see for instance Figure \ref{fig:potentials}), where the potential vanishes. This confirms that in full analogy with \cite{Konoplya:2004wg}, quasi-resonances of massive fields are forbidden in Weyl gravity not only for $k >0$, but also for $k=0$ and $\gamma<0$.

For \( r_0 = 3.14 \), the black hole transitions into a wormhole configuration. Nevertheless, the effective potential retains the same form in terms of the tortoise coordinate, allowing the same JWKB formalism to be applied to compute QNMs. From Tables 5 and 6, we observe that at \( r_0 = 5 \), the wormhole quasinormal frequencies significantly deviate from those of a black hole.

Our analysis reveals that the asymptotic value of the decay rate depends on the throat radius \( r_0 \) (as seen when comparing tables for \( \ell=0 \) and \( \ell=1 \)). However, we find that this value remains unchanged when varying the multipole number \( \ell \). This behavior arises because the effective potential includes a \( \mu^2 \) term that suppresses the centrifugal barrier, which otherwise scales as \( \ell (\ell + 1) r^{-2} \).

We can confirm the JWKB calculations with the time-domain integration. As for asymptotically de Sitter configurations the QNMs govern the decay of perturbations at all times \cite{Dyatlov:2010hq,Dyatlov:2011jd}, not only the fundamental mode but also the first  overtone of the Schwarzschild branch can be found with good accuracy \cite{Dubinsky:2024gwo}.  Typical time domain profiles can be seen in Figures \ref{fig:TD1} and \ref{fig:TD2}, where we see that the fundamental mode and the first overtone are extracted with a very good agreement with the JWKB data. An interesting behavior takes place for the wormhole case shown in Figure \ref{fig:TD2}, where we see that first an oscillatory tail with power-law envelope dominates in the signal and only later the exponential decay governed by QNMs takes place. A similar phenomenon was first observed for decay of a massive field in the wormhole background in \cite{Churilova:2019qph}. We can also see that the perturbation decay in time, indicating the stability of a massive scalar field.

\section{Gray-body factors}

\begin{figure}
\resizebox{\linewidth}{!}{\includegraphics{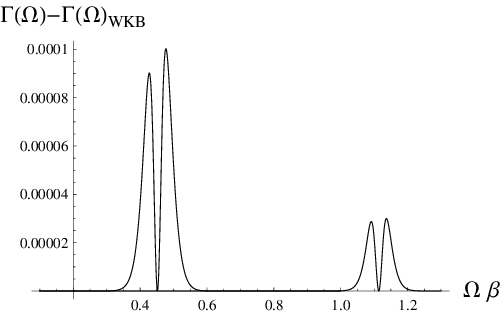}}
\caption{The difference between the gray-body factor calculated by the correspondence with QNMs and 6th order JWKB formula for $\ell=1$ scalar field for $\beta=1$  $\gamma=-0.3$ and various values of  $r_{0}=3.14$, $\mu =2$ (left) and $\mu=5$ (right).}\label{fig:GBF5}
\end{figure}
\begin{figure}
\resizebox{\linewidth}{!}{\includegraphics{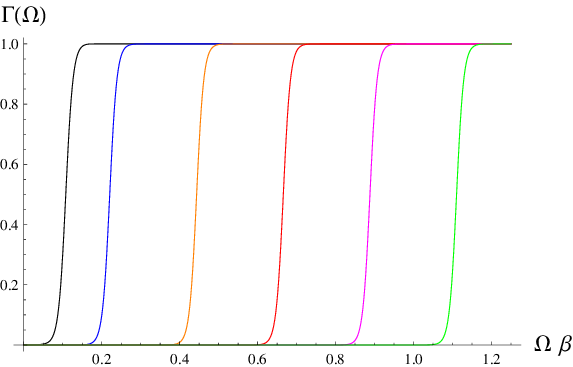}}
\caption{Gray-body factors for $\ell=0$ scalar field for the black hole $\beta=1$, $r_{0}=3.14$, $\gamma=-0.3$ for various values of mass $\mu =0.5$, $1$, $2$, $3$, $4$, $5$ (from left to right).}\label{fig:GBF1}
\end{figure}
\begin{figure}
\resizebox{\linewidth}{!}{\includegraphics{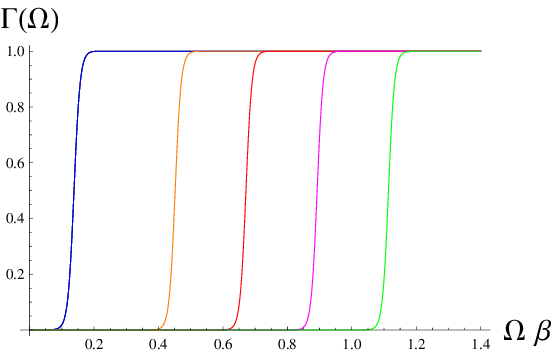}}
\caption{Gray-body factors for $\ell=1$ scalar field for the black hole $\beta=1$, $r_{0}=3.14$, $\gamma=-0.3$ for various values of mass $\mu =0.5$, $1$, $2$, $3$, $4$, $5$ (from left to right).}\label{fig:GBF2}
\end{figure}
\begin{figure}
\resizebox{\linewidth}{!}{\includegraphics{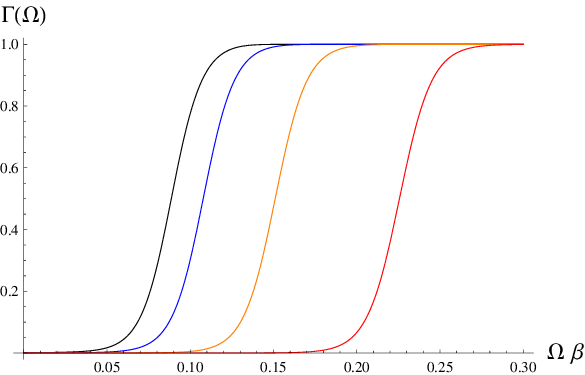}}
\caption{Gray-body factors for $\ell=0$ scalar field for $\beta=1$ $\mu=0.5$, $\gamma=-0.3$ and various values of  $r_{0}=0$, $3.14$, $5$, $6.28$ (from left to right).}\label{fig:GBF3}
\end{figure}
\begin{figure}
\resizebox{\linewidth}{!}{\includegraphics{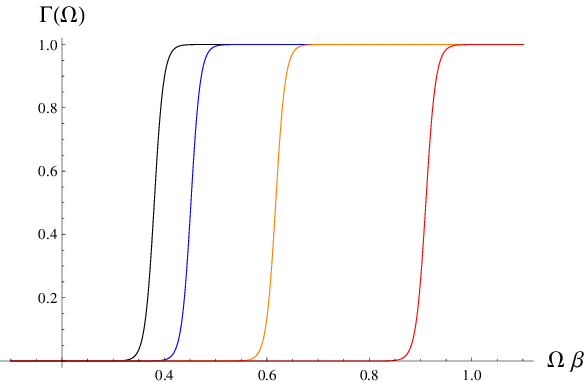}}
\caption{Gray-body factors for $\ell=1$ scalar field for $\beta=1$ $\mu=2$, $\gamma=-0.3$ and various values of  $r_{0}=0$, $3.14$, $5$, $6.28$ (from left to right).}\label{fig:GBF4}
\end{figure}

In the analysis of wave scattering in the vicinity of black holes and wormholes, an intriguing interplay arises between incoming radiation and the effective potential surrounding these compact objects. When a wave encounters the potential barrier, it undergoes partial reflection and partial transmission, leading to a phenomenon characterized by the so-called gray-body factors. These factors remain consistent regardless of whether the wave originates from the vicinity of the black hole’s event horizon (or, in the case of wormholes, from an asymptotic region in another universe) or whether it arrives from spatial infinity. This fundamental symmetry in black hole scattering imposes specific boundary conditions on the wave function, expressed as follows:

\begin{equation}
\begin{array}{rclcl}
\Psi &=& e^{-i\Omega r_*} + R e^{i\Omega r_*}, &\quad& r_* \to +\infty, \\
\Psi &=& T e^{-i\Omega r_*}, &\quad& r_* \to -\infty.
\end{array}
\end{equation}

Here, $\Psi$ denotes the wave function, $R$ represents the reflection coefficient, and $T$ is the transmission coefficient. The tortoise coordinate $r_*$ is defined such that the event horizon is mapped to $r_* \to -\infty$, while spatial infinity corresponds to $r_* \to +\infty$. A crucial distinction must be made between the real and continuous frequency $\Omega$, which characterizes wave scattering, and the complex and discrete QNM frequencies $\omega_n$, which define the characteristic oscillatory response of a perturbed black hole or wormhole.

While QNMs are extremely sensitive to even minor modifications of the near-horizon geometry \cite{Konoplya:2022pbc}, a contrasting property emerges in the behavior of gray-body factors, which exhibit remarkable robustness against such perturbations \cite{Rosato:2024arw,Oshita:2024fzf,Rosato:2025byu}. This relative stability makes gray-body factors a particularly valuable tool in studying the spectral properties of compact objects, as they provide an alternative means of probing the underlying spacetime structure without being significantly affected by microscopic deviations near the horizon.

In the context of black hole radiation and energy emission, the transmission coefficient $T$—commonly referred to as \textit{the gray-body factor}—plays a pivotal role. This factor determines the fraction of the wave that successfully overcomes the potential barrier, thereby contributing to the radiation escaping from the black hole.  The gray-body factor is defined as:
\begin{equation}
\Gamma_{\ell}(\Omega) = |T|^2 = 1 - |R|^2,
\end{equation}
where $\Gamma_{\ell}(\Omega)$ denotes the gray-body factor associated with a specific angular momentum mode $\ell$ and frequency $\Omega$. Since energy conservation dictates that the total incident energy must be redistributed between transmission and reflection, the sum of transmitted and reflected energy fractions necessarily equals unity. 

In this work, we leverage the well-established correspondence between QNMs and gray-body factors, as formulated in \cite{Konoplya:2024lir}.
\begin{equation}\label{eq:gbfactor}
\Gamma_{\ell}(\Omega) = \dfrac{1}{1+e^{2\pi i \K}},
\end{equation}
where
\begin{widetext}
\begin{eqnarray}\nonumber
&& i\K = \frac{\Omega^2 - \re{\omega_0}^2}{4\re{\omega_0}\im{\omega_0}}\Biggl(1 + \frac{(\re{\omega_0} - \re{\omega_1})^2}{32\im{\omega_0}^2}
-\frac{3\im{\omega_0} - \im{\omega_1}}{24\im{\omega_0}}\Biggr)
-\frac{\re{\omega_0} - \re{\omega_1}}{16\im{\omega_0}}
\\\nonumber&& -\frac{(\omega^2 - \re{\omega_0}^2)^2}{16\re{\omega_0}^3\im{\omega_0}}\left(1 + \frac{\re{\omega_0}(\re{\omega_0} - \re{\omega_1})}{4\im{\omega_0}^2}\right)
+\frac{(\omega^2 - \re{\omega_0}^2)^3}{32\re{\omega_0}^5\im{\omega_0}}\Biggl(1 + \frac{\re{\omega_0}(\re{\omega_0} - \re{\omega_1})}{4\im{\omega_0}^2}
\\\nonumber&&\qquad + \re{\omega_0}^2\Biggl(\frac{(\re{\omega_0} - \re{\omega_1})^2}{16\im{\omega_0}^4}
-\frac{3\im{\omega_0} - \im{\omega_1}}{12\im{\omega_0}}\Biggr)\Biggr) + \Order{\frac{1}{\ell^3}},
\label{eq:gbsecondorder}
\end{eqnarray}
\end{widetext}
and $\omega_0$ and $\omega_1$ are the fundamental mode and the first overtone of the Schwarzschild branch.
Alternatively, one can use the JWKB formula \ref{eq:gbfactor}, where $\K$ can be found via the 6th order JWKB approach that requires a knowledge of the effective potential and its derivatives up to the 12th order in the maximum \cite{Konoplya:2019hlu}.

Remarkably, this correspondence has been shown to hold with extraordinary precision in scenarios involving massive fields propagating in asymptotically de Sitter backgrounds \cite{Malik:2024cgb}, which is precisely the case under consideration here. 
We confirm this observation via comparison of the gray-body factors calculated via the correspondence with those found by the 6th order JWKB formula (see Figure \ref{fig:GBF5}). There one can see that the difference is a tiny fraction of one percent or smaller.  Thus, this correspondence provides a powerful analytical and numerical framework for investigating the transmission properties of massive perturbations in black hole and wormhole spacetimes.

When mass of the field $\mu$ is increased, the gray-body factors are strongly suppressed, so that the regime of almost complete reflection is prolonged to larger frequencies as shown in Figures \ref{fig:GBF1} and \ref{fig:GBF2}.

It is worth of noticing that for massless fields the gray-body factors were calculated in \cite{Konoplya:2025mvj} only for the Schwarzschild-like  Mannheim-Kazanas solution, which corresponds to $r_{0}$ case. Here we calculate gray-body factors for various $r_{0}$ including thereby non-Schwarzschild black holes and traversable wormholes. From Figures \ref{fig:GBF3} and \ref{fig:GBF4} we can see that the gray-body factor are smaller for larger deviations from the Schwarzschild limit $r_{0}=0$. 

It is important to note that the correspondence between QNMs and gray-body factors holds under two specific conditions: 
\begin{enumerate}
    \item Only the Schwarzschild branch of QNMs allows for an accurate reproduction of the gray-body factors.
    \item The correspondence remains valid only when the effective potential exhibits a standard barrier shape with a single peak. 
\end{enumerate}
In scenarios (which do not arise in our configurations) where the transition from a black hole to a wormhole results in the emergence of a double-peaked potential, the correspondence ceases to be applicable \cite{Konoplya:2025hgp}.

\section{Conclusions}\label{sec:conclusions}

The spectral characteristics of massive fields around compact objects, such as black holes and wormholes, in the regime
\begin{equation}
\frac{\mu M}{m_P} \gg 1
\end{equation}
play a crucial role, as this regime encompasses the radiation of massive particles from the Standard Model when the compact object's mass exceeds approximately $10^{22}$ kg. This corresponds to a microscopic black hole of around $\sim 10^{-5}$ m. At the same time, the opposite regime of ultralight particles is expected to contribute to radiation at extremely long wavelengths \cite{Konoplya:2023fmh}, which have recently been detected in the Time Pulsar Array experiment \cite{NANOGrav:2023gor}.  

In this work, we have investigated the QNMs and gray-body factors of a massive scalar field in the background of three types of compact objects: Schwarzschild-like black holes, non-Schwarzschild black holes, and traversable wormholes, which emerge as exact solutions in Weyl gravity \cite{Mannheim:1988dj,Jizba:2024owd}. Our findings reveal that the spectrum of massive fields differs qualitatively from that of massless fields, leading to significantly longer-lived modes while avoiding the regime of arbitrarily long-lived modes observed in asymptotically flat black holes and wormholes \cite{Ohashi:2004wr,Konoplya:2004wg}.  

In the time domain, perturbations exhibit a distinctive decay pattern: upon transitioning to the wormhole state, oscillatory tails with a power-law envelope precede the quasinormal ringing, which decays exponentially. The gray-body factors are significantly suppressed as the parameter $r_{0}$ increases, enabling a clear distinction between Schwarzschild-like black holes and traversable wormholes in Weyl gravity. Furthermore, we have observed that the correspondence between QNMs and gray-body factors holds with remarkable accuracy in our case.  

Future extensions of this work could involve studying massive fields of other spins, such as Proca and Dirac fields, to explore their unique spectral properties in similar backgrounds.

\section*{Acknowledgments}
The author is grateful to Excellence Project PřF UHK 2205/2025-2026 for the financial support.

\bibliography{Bibliography}
\end{document}